%% file: main.tex
\documentclass{article}

\usepackage{arxiv}

\usepackage[utf8]{inputenc} 
\usepackage[T1]{fontenc}    
\usepackage{hyperref}       
\usepackage{url}            
\usepackage{booktabs}       
\usepackage{amsfonts}       
\usepackage{nicefrac}       
\usepackage{microtype}      
\usepackage{lipsum}		
\usepackage{graphicx}
\usepackage{natbib}
\usepackage{doi}
\usepackage{moreverb}
\usepackage{multirow}
\usepackage{multicol}
\usepackage{comment}
\usepackage{rotating}

\usepackage{acro}
\usepackage{bm}
\usepackage{algorithm} 
\usepackage{algorithmicx} 
\usepackage{algpseudocode}  

\newcommand\BibTeX{{\rmfamily B\kern-.05em \textsc{i\kern-.025em b}\kern-.08em
T\kern-.1667em\lower.7ex\hbox{E}\kern-.125emX}}

\usepackage{amsmath}
\usepackage{amssymb}
\usepackage{enumitem}
\usepackage{pdfpages}
\usepackage{blkarray}
\usepackage{mathtools}
\usepackage{natbib}
\usepackage{url}
\usepackage{graphicx,subfigure}
\usepackage{epstopdf} 
\usepackage{array}
\usepackage{float}
\usepackage{ragged2e}

\DeclareMathOperator*{\argmax}{arg\,max}
\DeclareMathOperator*{\argmin}{arg\,min}

\usepackage{color}
\usepackage{comment}
\usepackage{ulem}
\usepackage{soul}

\title{Extensions of Heterogeneity in Integration and Prediction (HIP) with {R Shiny} Application}

\author{ 
	{\hspace{1mm}Jessica Butts, PhD} \\
	Division of Biostatistics\\
	University of Minnesota Twin Cities\\
        \And 
        {\hspace{1mm}Christine Wendt, MD} \\
	Division of Pulmonary, Allergy and Critical Care\\
	University of Minnesota Twin Cities\\
        \And 
        {\hspace{1mm}Russel  Bowler, MD, PhD} \\
	Division of Pulmonary, Critical Care and Sleep Medicine\\
        Department of Medicine\\
	National Jewish Health\\
        \And 
        {\hspace{1mm}Craig P. Hersh, MD} \\
	Channing Division of Network Medicine\\
        Brigham and Women's Hospital\\
	Harvard Medical School\\
        \And 
        {\hspace{1mm}Qi Long, PhD} \\
	Department of Biostatistics, Epidemiology and Informatics\\
 Perelman School of Medicine\\
	University of Pennsylvania\\
  \And 
        \And 
        {\hspace{1mm}Lynn Eberly, PhD} \\
	Division of Biostatistics\\
	University of Minnesota Twin Cities\\
  \And 
        \href{https://orcid.org/0000-0001-9593-4778}{\includegraphics[scale=0.06]{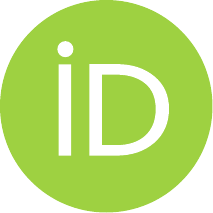}\hspace{1mm}Sandra E. Safo}\thanks{Corresponding Author: Sandra E. Safo, www.sandraesafo.com} \\
	Division of Biostatistics\\
	University of Minnesota Twin Cities\\
	\texttt{ssafo@umn.edu} \\
}

\renewcommand{\shorttitle}{HIP with Shiny R}

\hypersetup{
pdftitle={A template for the arxiv style},
pdfsubject={q-bio.NC, q-bio.QM},
pdfauthor={David S.~Hippocampus, Elias D.~Striatum},
pdfkeywords={First keyword, Second keyword, More},
}

\begin{document}
\maketitle

\begin{abstract}
Multiple data views measured on the same set of participants is becoming more common and has the potential to deepen our understanding of many complex diseases by analyzing these different views simultaneously. Equally important, many of these complex diseases show evidence of subgroup heterogeneity (e.g., by sex or race). HIP (Heterogeneity in Integration and Prediction) is among the first methods proposed to integrate multiple data views while also accounting for subgroup heterogeneity to identify common and subgroup-specific markers of a particular disease. However, HIP is applicable to continuous outcomes and requires programming expertise by the user.  Here we propose extensions to HIP that accommodate multi-class, Poisson, and Zero-Inflated Poisson outcomes while retaining the benefits of HIP. Additionally, we introduce an R Shiny application, accessible on shinyapps.io at \url{https://multi-viewlearn.shinyapps.io/HIP_ShinyApp/},
that provides an interface with the Python implementation of HIP to allow more researchers to use the method anywhere and on any device. We applied HIP to identify genes and proteins common and specific to males and females that are associated with exacerbation frequency. Although some of the identified genes and proteins show evidence of a relationship with chronic obstructive pulmonary disease (COPD) in existing literature, others may be candidates for future research investigating their relationship with COPD. We demonstrate the use of the Shiny application with a publicly available data. An R-package for HIP would be made available at \url{https://github.com/lasandrall/HIP}. 
\end{abstract}

\keywords{Multi-view data; Subgroup Heterogeneity; Integrative Analysis; COPD; Multimodal; Multi-omics}

\section{Introduction}\label{ext:sec:intro}

Chronic obstructive pulmonary disease (COPD) is a chronic disease affecting the lungs and airways in almost 4\% of the global population in 2017 \citep{soriano2020prevalence}. Cigarette smoking is a known risk factor for COPD, but fewer than 50\% of heavy smokers develop COPD \citep{agusti2023global}. There are also many genetic and environmental factors influencing risk \citep{silverman2020review, hardin2014chronic, hu2010risk, chung2008multifaceted}. COPD research is further complicated by the subgroup heterogeneity that exists between males and females. In a meta-analysis, female smokers had a faster annual decline in forced expiratory volume in one second (FEV$_{1}$) even if they smoked fewer cigarettes \citep{gan2006female}. Another study found that women smokers generally had higher airway wall thickness (AWT) compared to male smokers \citep{kim2011gender}. Additionally, researchers found that women experienced increased risk of hospitalization for COPD compared to males even when controlling for smoking \citep{prescott1997gender}.

The Genetic Epidemiology of COPD (COPDGene) Study \citep{regan2011genetic} was designed to understand genetic factors related to the development of COPD. The study collected genetic and proteomic data on a subset of participants at the Phase 2 (P2) study visit. Given the availability of multi-view data (e.g., genomic and proteomic data) and the known subgroup heterogeneity between males and females, the application of an integrative analysis method accounting for subgroup heterogeneity to the COPDGene Study data offers the opportunity to gain new insights into COPD. Butts et al. \cite{butts_2023_HIP} recently proposed a method called HIP, short for Heterogeneity in Integration and Prediction, for integrating data from multiple sources and simultaneously predicting a \textit{continuous} outcome while accounting for subgroup heterogeneity. HIP allows to identify \textit{common} and \textit{subgroup-specific} variables contributing most to the overall association among the views and the variation in the outcome.  HIP was used to  investigate  airway wall thickness (AWT) as a proxy for COPD severity and the authors demonstrated that HIP was capable of identifying genes and proteins common and specific to males and females that were predictive of AWT. However, AWT is not the only way to characterize the effects of COPD, and one may be interested in different types of outcomes that are not continuous. 

One such outcome of interest is the number of COPD exacerbations, generally defined as an acute worsening of symptoms that require a change in treatment; these symptoms can include cough, wheezing, dyspnea, chest tightness, and decreased exercise tolerance. While exacerbations vary in severity, severe cases may need to be hospitalized and put on a ventilator; patients who require ICU treatment have a 43-46\% risk of death within a year of the hospitalization \citep{evenson2010management}. As such the number of exacerbations experienced by patients is a clinically meaningful outcome. Additionally, the TORCH (Towards a Revolution in COPD Health) study found that females had a rate of exacerbations that was 25\% higher than males during their 3-year follow-up \citep{celli2011sex}. This emphasizes not only the importance of looking at exacerbation frequency but also accounting for differences in sex. Because exacerbation frequency is a Poisson or rate variable, it is not compatible with the originally proposed HIP \cite{butts_2023_HIP}.

\begin{figure}[htbp]
    \centering
    \includegraphics[width=0.7\textwidth]{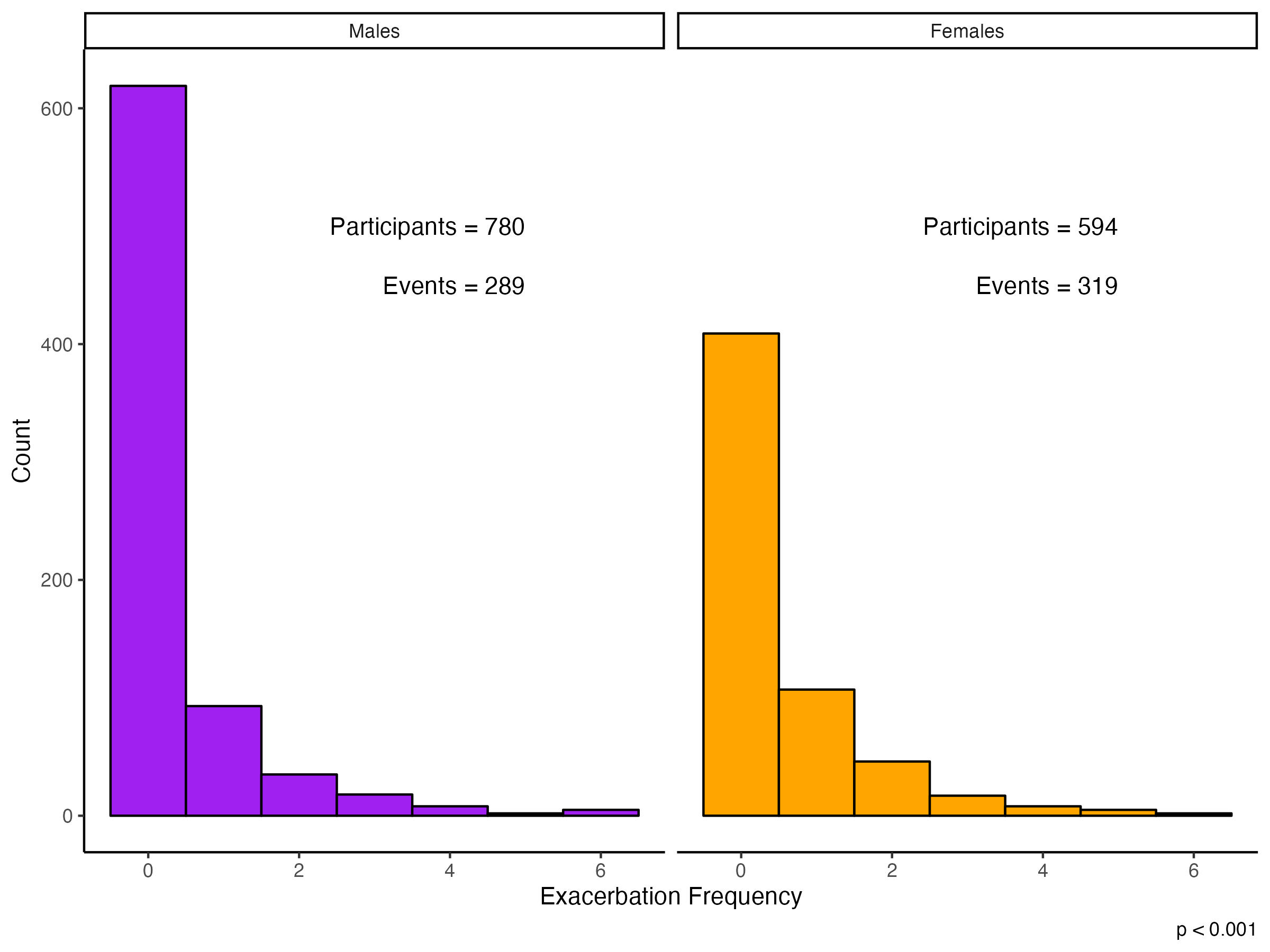}
    \caption{{\bf Distribution of Exacerbation Frequency for Males and Females at COPDGene Study Visit P2.} The figure includes data from the P2 visit of the COPDGene Study for the subset of subjects included in our analysis. There is a statistically significant difference in the number of exacerbations experienced per person-year between males and females.}\label{fig:Exacerbation_Hist}
\end{figure}

Our goal is to deepen our investigation into the molecular underpinnings of sex differences in COPD mechanisms using data from the COPDGene Study \citep{regan2011genetic} by identifying genes and proteins common and specific to males and females related to exacerbation frequency. For this analysis, we included COPDGene Study \citep{regan2011genetic} participants with COPD (defined as GOLD stage $\geq 1$) at P2 and with genomic, proteomic, and selected clinical covariates (age, BMI, race, pack-years, FEV$_1$\%, AWT, and \% emphysema) available. Figure \ref{fig:Exacerbation_Hist} shows the distributions of number of exacerbations in the past year for males and females in this subsample. First, we note a statistically significant difference in the number of events per person-year between males and females. Second, both males and females had several participants with zero exacerbations suggesting that a zero-inflated Poisson (ZIP) outcome may better fit this data. 

To address our research goal, the simplest approach we could consider would be to use a penalized regression method such as the Elastic Net \citep{enet} or Lasso \citep{lasso}. The \emph{glmnet} package \citep{glmnet} can implement the Lasso and Elastic Net on a binary or Poisson outcome, but these methods neither perform integrative analysis nor account for subgroup heterogeneity; they also cannot accommodate a ZIP outcome. These approaches would require concatenating data views as they can only accept a single data view; they would also require running separate analyses for each subgroup to allow for subgroup heterogeneity. These approaches are easily implemented but are not ideal for our research goal where we want to integrate data from multiple sources and associate these data with a ZIP outcome  while accounting for subgroup heterogeneity.

There are several integrative analysis methods we could consider for this analysis, but to the best of our knowledge, none of them account for subgroup heterogeneity, and few can accommodate a ZIP outcome. For example, Canonical Variate Regression (CVR) \citep{luo2016canonical} is a one-step  method for simultaneously associating data from multiple sources and predicting a clinical outcome.  CVR can accommodate continuous, binary or Poisson outcomes but not a ZIP outcome. Sparse Integrative Discriminant Analysis (SIDA) \citep{SIDA} is a one-step method for joint association and classification of data from multiple sources, but it is only applicable to classification problems (i.e., binary or multi-class outcome). The aforementioned methods are one-step in that the problem of associating the multiple views are coupled with the problem of predicting an outcome. We could use a two-step method to first model the associations between views and then model the Poisson or ZIP response using information from the first step. For example, we could perform canonical correlation analysis (CCA) using SELP (Sparse Estimation through Linear Programming) \citep{safo_2018_selpcca} and then use the canonical variates as predictors in a Poisson or ZIP regression model. However, these all still fail to account for subgroup heterogeneity. To use any of these integrative analysis methods, we would have to either (a) concatenate the subgroups which ignores any potential subgroup heterogeneity or (b) run a separate analysis for each subgroup which limits power  especially in a high-dimensional data setting where the sample size is typically less than the number of variables.

Alternatively, we could consider methods that account for subgroup heterogeneity, but to the best of our knowledge, none of them perform integrative analysis. One example is the Joint Lasso \citep{dondelinger2018joint}, but this method is unable to accommodate a binary, Poisson, or ZIP outcome. More generally, methods that account for subgroup heterogeneity but do not perform integrative analysis would require either (a) concatenating data views within each subgroup which fails to model the associations between data views or (b) considering each view separately which fails to fully utilize the multi-view data and requires combining results post hoc.

With all of these existing methods, the only way to get subgroup-specific information is to fit the model separately on each subgroup, but this can greatly reduce the sample size and thus power to detect effects. Alternatively, if these methods are run on the combined subgroup data, the presence of any potential subgroup heterogeneity will be ignored. Existing methods do not fully utilize the available COPDGene Study \citep{regan2011genetic} data to address our interest in identifying genes and proteins common and specific to males and females related to exacerbation frequency (modeled as a ZIP outcome). Thus, we propose extensions of HIP that can accommodate multi-class, Poisson, and ZIP outcomes while preserving the existing benefits of HIP (joint association and prediction method for integrative analysis accounting for subgroup heterogeneity and feature ranking) in order to identify common and subgroup-specific features predictive of an outcome. Further, since HIP requires Python programming expertise which is limiting, we develop a web application using the R Shiny framework and hosted on shinyapps.io, allowing HIP to be used anywhere, on any device, and by users with limited programming expertise. We believe that this will increase the widespread adoption of HIP by biomedical researchers interested in integrative analysis methods that also account for subgroup differences. 

The remainder of this paper is structured as follows. In Section \ref{ext:sec:methods}, we present the proposed methods to extend HIP. In Section \ref{ext:sec:alg}, we describe the algorithmic implementation of this extension. In Section \ref{ext:sec:sims}, we describe the design and results of simulations assessing the performance of HIP in comparison with existing methods. In Section \ref{ext:sec:rda}, we apply HIP to data from the COPDGene Study \citep{regan2011genetic} to examine the relationship of genes and proteins with exacerbation frequency. Section \ref{sec:shiny} introduces an R Shiny \citep{r_shiny} application that provides a graphical interface to the underlying Python implementation of HIP. We conclude with discussion of limitations and future work in Section \ref{ext:sec:conc}.

\section{Methods}\label{ext:sec:methods}

\subsection{Notation and Problem}\label{ext:sec:methods:notation}
Consider the scenario where we have $D$ data views (e.g., genomics, proteomics, clinical) measured on the same set of $N$ subjects. Each view has $p_d$ variables and $S$ subgroups (known a priori). Each subgroup has sample size $n_s$. For our application, $S=2$ for biological sex (males and females). The total number of samples is $N=\sum_{s=1}^{S}n_s$. The data matrix for subgroup $s$ is denoted by $\bm X^{d,s} \in \mathcal{R}^{n_s \times p_d}$ where rows are for samples and columns are for variables. We define the outcome for each subgroup as $\bm Y^s$. For a multi-class outcome, $\bm Y^s \in \mathcal{R}^{n_s \times m}$ with $m$ being the number of classes, is an indicator matrix where each row has a one in the column corresponding to the class of the observation and a 0 in all other columns. For a Poisson or ZIP outcome, $\bm Y^s \in \mathcal{R}^{n_s \times 2}$ where the first column is the observed counts and the second column is an offset. If no offset is provided, we add a column of ones as the offset which does not affect estimation. 

\subsection{Existing HIP Framework}
Since we extend the original HIP to accommodate different types of outcomes  beyond continuous outcome(s), we briefly introduce HIP for completeness sake. HIP  simultaneously associates data from multiple views, predicts an outcome, and ranks or identifies common and subgroup-specific variables contributing to the overall dependency structure and the variation in an outcome  by minimizing an objective function that is a sum of three terms: 
association term, hierarchical penalty term, and a prediction term. Consider the association term. HIP assumes that  each view $\bm X^{d,s}$ can be approximated by a product of low-rank view-independent matrix, $\bm Z^s  \in \mathcal{R}^{n_s \times K}$ and view-specific loadings $\bm B^{d,s} \in \mathcal{R}^{p_d \times K}$.
That is, HIP assumes that $\bm X^{d,s} = \bm Z^s {\bm B^{d,s}}^T + \bm{E^{d,s}}$ where $\bm Z^s$ are latent scores describing association across views and is shared by all views for subgroup $s$, $\bm B^{d,s} \in \mathcal{R}^{p_d \times K}$ are view and subgroup-specific variable loadings, and $\bm E^{d,s}$ are errors as a result of approximating $\bm X^{d,s}$ with $\bm Z^s {\bm B^{d,s}}^T$. Here, $K$ is the number of latent components, which is typically $K \ll p_d$, resulting in data dimensionality reduction. Then, $\bm B^{d,s}$ and $\bm Z^s$ are estimated to minimize the difference between the observed and estimated data via the loss function:   

\begin{equation}
    F(\bm X^{d,s}, \bm Z^s, \bm B^{d,s})= \|\bm X^{d,s} - \bm Z^s {\bm B^{d,s}}^T\|_F^2,
    \label{eq:association}
\end{equation}
where $\| \cdot \|_F$ is the Frobenius norm. 
We now consider the hierarchical penalty term which allows for identifying common and subgroup-specific variables. The hierarchical penalty implemented in HIP is a modified version of the penalty proposed in the Meta Lasso \citep{li2014meta}. For common and subgroup-specific variable selection, HIP decomposes $\bm B^{d,s}$ into the element-wise product of $\bm G^d$ and $\bm \Xi^{d,s}$, i.e. $\bm B^{d,s} =\bm G^d \cdot \bm \Xi^{d,s}$. Here, $\bm G^d$ allows us to model common effects across subgroups while $\bm \Xi^{d,s}$ allows us to model  heterogeneity between subgroups. Since $\bm G^d$ is common to all subgroups, its estimation allows us to borrow strength from all subgroups and improves power in estimation especially in high-dimensional settings where the number of samples is smaller than the number of variables. Of note, in this reparameterization, exact values of $\bm G^d$ and $\bm \Xi^{d,s}$ are not identifiable, but also are  not directly needed for variable ranking since variable ranking is based on $\bm B^{d,s}$. HIP imposes a block $L_{2,1}$ penalty on both $\bm G^d$ and $\bm \Xi^{d,s}$ which encourages selection of a variable in all $K$ components or none of the components. Specifically, the hierarchical penalty term proposed in HIP is:
\begin{equation}
    \sum_{d=1}^D \sum_{s=1}^S \mathcal{J}(\bm B^{d,s}) = \lambda_G \sum_{d=1}^D \gamma_d \sum_{l=1}^{p_d} \| {\bm g_l^d}\|_2 + \lambda_\xi \sum_{d=1}^D \gamma_d  \sum_{s=1}^S\sum_{l=1}^{p_d} \| \bm \xi_l^{d,s}\|_2, 
    \label{eq:penalty}
\end{equation}
where $\bm g_l^d$ is a vector of length $K$ in the $l$th row of $\bm G^d $ and $\bm \xi_l^{d,s}$ is a vector of length $K$ in the $l$th row of $\bm \Xi^{d,s}$. 
The strength of the penalty on $\bm G^d$ and $\bm \Xi^{d,s}$ is controlled by  hyperparameters $\lambda_G > 0 $ and $\lambda_\xi >0 $ respectively. The $\gamma_d$ is a user-specified indicator of whether to penalize view $d$, i.e. whether to select variables in view $d$. This term is useful to force the inclusion of covariates believed to affect the outcome in the integrative analysis model. 

We now consider the prediction term which relates the outcome to the association term via $\bm Z^s$ using the loss function:
$F(\bm Y^s, \bm Z^s, \bm \Theta, \beta_0) = \| \bm Y^s -  \mathbf{\mathcal{J}}_{n_s} \beta_0^T-  \bm Z^s \bm \Theta^T\|_F^2$, where $\Theta \in \mathcal{R}^{q \times K}$  is a matrix of regression coefficients, $\mathbf{\mathcal{J}}_{n_s}$ is a vector of ones of size $n_s$ and $\beta_0^T$ is a vector of length $q$ for the intercept. $\bm \Theta$ is common and not subgroup-dependent. This allows the outcome data for all subgroups to be used in estimating the parameters and can improve the overall prediction of the outcome. Of note, for single continuous outcome, $q =1$, and for multiple continuous outcome, $q >1$. The presence of $\bm Z^s$ in both the prediction and association terms is what allows HIP to be a one-step method. In other words, HIP simultaneously models the association between multiple views and predicts an outcome, compared to two-step methods that separate the association and prediction problems which may result in the $\bm Z^s$ not being clinically meaningful.  Since the two steps are combined in HIP, the estimation of the low-dimensional view-independent components $\bm Z^s$ is guided by an outcome and therefore $\bm Z^s$ is naturally endowed with prediction capabilities. Put together, HIP solves the following optimization problem to estimate the subgroup-specific view-independent components $\bm Z^s$, view-specific common variables $\bm G^d$, view- and subgroup-specific variables $\bm \Xi^{d,s}$ and regression coefficients $\bm \Theta$ that associate multiple views, predicts a continuous outcome(s) and identifies common- and subgroup-specific variables: 
\begin{align} 
( \hat{\bm B}^{d,s}, \hat{\bm Z}^s, \hat{\bm \Theta}, \hat{\beta}_0) &= \underset{\bm B^{d,s}, \bm Z^s, \bm \Theta, \beta_0}{\text{min}} \sum_{s=1}^S F(\bm Y^s, \bm Z^s, \bm \Theta, \beta_0) + \sum_{d=1}^D \sum_{s=1}^S F(\bm X^{d,s}, \bm Z^s, \bm B^{d,s}) +  
\sum_{d=1}^D\sum_{s=1}^S \mathcal{J}(\bm B^{d,s})
\label{objective}
\end{align}
Because the prediction term relies on the type of outcome, in the current work, we redefine $F(\bm Y^s, \bm Z^s, \bm \Theta, \beta_0)$ based on the type of outcome. We define $F(\bm Y^s, \bm Z^s, \bm \Theta, \beta_0)$ for multi-class, Poisson, and ZIP outcomes below.

\subsection{Beyond Gaussian outcome(s): Modification of HIP Prediction Term}

\subsubsection{Multi-class Outcome}
Similar to the continuous outcome, we relate the shared component $\bm Z^s$ with the multi-class outcome.  For this purpose, we use the cross-entropy loss function: 
$    F(\bm Y^s, \bm Z^s, \bm \Theta, \beta_0) = -\sum_{i=1}^{n_s} \sum_{j=1}^m y_{ij}^s \log(a_{ij}^s)$,
where $a_{ij}^s = \frac{\exp\{w_{ij}^s\}}{\sum_{j=1}^m \exp\{w_{ij}^s\}}$ is the \textit{softmax} function generalizing logistic regression from binary to multi-class problems. Here, $\bm W^s = \mathbf{\mathcal{J}}_{n_s}\beta_0 + \bm Z^s \bm \Theta$ represents scores, $\mathbf{\mathcal{J}}_{n_s}$ is an $n_s \times 1$ matrix of ones, $\bm \Theta \in \mathcal{R}^{K \times m}$, and $w_{ij}^s$ is the $ij$th entry in $\bm W^s$. 
The \textit{softmax} function forces the sum of each row in $\bm W^s$ to be 1 so that each entry in the row represents a probability of case $i$ belonging to class $j$.

\subsubsection{Poisson Outcome}
For a Poisson outcome, we define the prediction term: 
    $F(\bm y^s, \bm Z^s, \bm \Theta, \beta_0) = \sum_{i=1}^{n_s} - y_i^s[\log(t_i^s)+\beta_0+ \bm Z_i^s \bm \Theta] + t_i^s\exp(\beta_0 + \bm Z_i^s \bm \Theta) + \log(y_i^s!)$,
which is based on the negative log-likelihood for a Poisson regression with an offset using a log link. Here, $y_i^s$ is the outcome for the $i$th subject in subgroup $s$, $t_i^s$ is the offset for the $i$th subject in subgroup $s$, and $\bm Z_i^s$ is the $i$th row of the $\bm Z^s$ matrix.  Here $\bm \Theta \in \mathcal{R}^{K \times 1}$ and $\beta_0 \in \mathcal{R}^{1 \times 1}$.

\subsubsection{Zero-Inflated Poisson Outcome}
As mentioned in Section \ref{ext:sec:intro}, the motivating data set has a ZIP outcome rather than a true Poisson outcome. In order to accommodate this, we add an additional loss function to the code based on Lambert  \cite{lambert1992zip}. In this model, the observed outcome is from the zero state with probability $\tau$ and a Poisson random variable with probability $1-\tau$. We assume that covariates are only related to the Poisson mean ($\lambda$) and not $\tau$. Further, we assume that there is no relationship between $\lambda$ and $\tau$. Thus, $\log(\lambda_i) = \log(t_i) + \beta_0 + \bm Z_i^s \bm \Theta$. This assumption is a simplifying assumption, so while we could miss modeling a relationship that is there, it is not enforcing or requiring any specific constraints. Using this distribution, we use the negative log-likelihood to define the loss function as: 
\begin{align}
    F(\bm Y^s, \bm Z^s, \bm \Theta, \beta_0) = &\sum_{y_i^s=0} -\log(\exp[\tau] + \exp[-t_i^s \exp(\beta_0 + \bm Z_i^s \bm \Theta)])  \nonumber \\
    - &\sum_{y_i^s > 0} \big( y_i^s[\log(t_i^s)+ \beta_0 + \bm Z_i^s \bm \Theta] + t_i^s\exp[\beta_0 + \bm Z_i^s \bm \Theta] \big) \nonumber \\
    + &\sum_{i=1}^{n_s} \big( \log(y_i^s!) +  \log[1+\exp(\tau)] \big).
    \label{eq:zip}
\end{align}
This does require estimation of the additional parameter $\tau$ which is described in Section \ref{ext:sec:alg}. Again $\bm \Theta \in \mathcal{R}^{K \times 1}$ and $\beta_0 \in \mathcal{R}^{1 \times 1}$.
\subsection{Prediction}\label{prediction}
Suppose we have a test data, $\bm X^{d,s}_{test}$, $s=1,\ldots, S$, $d=1,\ldots, D$. Our goal in this section is to use the estimated optimization parameters $\widehat{\bm Z^s}$, $\widehat{\bm \Theta}$, and $\widehat{\beta}_0$  and the test data $\bm X^{d,s}_{test}$ to predict the test outcome  $\widehat{\bm Y}^s$ for subgroup $s$. We first estimate the test shared component for subgroup $s$, $\widehat{\bm Z}^s_{pred}$,  by solving the optimization problem: 
\begin{equation}\label{eq:pred}
    \widehat{\bm Z}^s_{pred}= \underset{ \bm Z^s}{\text{min}} \sum_{d=1}^D \sum_{s=1}^S F(\bm X^{d,s}_{test}, \bm Z^s, \hat{\bm B}^{d,s}) = \underset{ \bm Z^s}{\text{min}}\sum_{d=1}^D \sum_{s=1}^S\|\bm X^{d,s}_{test} - \bm Z^s \hat{\bm B}^{{d,s}^T}\|_F^2.
\end{equation}
Let $\bm X_{cat}^s$ be an $n_s \times \{p_1 + \cdots + p_d\}$ matrix that concatenates all $D$ views for subgroup $s$, i.e., $X_{cat}^s=[ \bm X^{1,s}_{test},\cdots,\bm X^{D,s}_{test}]$. Similarly, let $\widehat{\bm B}_{cat}= [\widehat {\bm B}^{1,s},\cdots, \hat{\bm B}^{D,s}]$ be a $\{p_1 + \cdots + p_D \} \times K$ matrix of variable coefficients. Then the solution to the optimization problem \eqref{eq:pred} is given as: $\widehat{\bm Z}^s_{pred} = \bm X_{cat}^s {\widehat{\bm B}_{cat}^s}({\widehat{\bm B}_{cat}^{s^T}} {\widehat{\bm B}_{cat}^s})^{-1}$ for $s=1,\ldots,S$. Given the predicted $\widehat{\bm Z}^s_{pred}$, we predict the test outcome as follows: 

\begin{equation*}
    \widehat{y}^s_{i} =
\begin{cases}
    \argmax_{j} \frac{\exp\{\widehat{w}_{ij}^s\}}{\sum_{j=1}^m \exp\{\widehat{w}_{ij}^s\}} & \text{Multi-class}\\
    t_i^s \exp(\widehat{\beta}_0 + \widehat{\bm Z}_{{pred}_i}^s \widehat{\bm \Theta}) & \text{Poisson} \\
    (1-\widehat{\tau}) t_i^s \exp(\widehat{\beta}_0 + \widehat{\bm Z}_{{pred}_i}^s \widehat{\bm \Theta}) & \text{ZIP},
\end{cases}
\label{ext:eq:pred}
\end{equation*}
where $\widehat{\bm Z}_{{pred}_i}^s$ is the $i$th row of $\widehat{\bm Z}_{pred}^s$, $\widehat{w}_{ij}^s$ is the $ij$th element of $\widehat{\bm W}^s = \mathbf{\mathcal{J}}_{n_s}\widehat{\beta}_0 + \widehat{\bm Z}_{pred}^s \widehat{\bm \Theta}$.

\section{Algorithm}\label{ext:sec:alg}
\subsection{Optimizing Parameters and Ranking Variables}
We use alternating minimization algorithm to solve optimization problem \eqref{objective}. We initialize the entries of ${\bm Z^s}^{(0)}$ by randomly sampling from a $U(0.9, 1.1)$ distribution. We initialize the entries of $\bm G^{d^{(0)}}$ for $d = 1, ..., D$,  $\bm \Theta^{(0)}$, and $\beta_0^{(0)}$ with ones. We initialize $\bm \Xi^{{d,s}^{(0)}}$ as  $\bm \Xi^{{d,s}^{(0)}} = [({\bm Z^{s^{(0)}}}^T \bm Z^{s^{(0)}})^{-1} {\bm Z^{s^{(0)}}}^T \bm X_{train}^{d,s}]^T$.

We estimate $\hat{\bm Z^s}^{(t)}$ at iteration $t$ by optimizing the following problem  using gradient descent with gradients calculated using PyTorch \cite{PyTorch}:
\begin{equation}\label{eq:zopt}
\hat{\bm Z^s}^{(t)} = \underset{\bm Z^s}{\text{min}} \sum_{s=1}^S F(\bm Y^s, \bm Z^s, \bm \Theta^{(t-1)}, \beta_0^{(t-1)}) + \sum_{d=1}^D \sum_{s=1}^S \|\bm X^{d,s} - \bm Z^s \bm B^{{{d,s}^{(t-1)}}^T}\|_F^2,
\end{equation}
where $F(\bm Y^s, \bm Z^s, \bm \Theta^{(t-1)}, \beta_0^{(t-1)})$ depends on the type of outcome. We use FISTA (fast iterative shrinkage-thresholding algorithm) with backtracking \cite{FISTA} to speed up convergence and select an appropriate step size. For a ZIP outcome, further consideration has to be given to the additional parameter, $\tau$. 
The parameter $\tau$ is the probability that a given observation is in the zero state. Note that this differs from $P(y_i^s = 0) = \tau + (1 - \tau)e^{-\lambda_i}$ as $P(y_i^s = 0)$ includes the probability that is in the zero state plus the probability of a zero from a Poisson distribution with mean $\lambda_i$. We initialize $\tau$ using the observed proportion of excess $0$s beyond the proportion predicted by the Poisson model across all observations \citep{lambert1992zip} shown in \eqref{eq:excess0}. Specifically, the first term is the observed proportion of zeros, and the second term is the $P(y_i^s = 0 | \lambda_i = \beta_0 + \bm Z_i^s \bm \Theta)$ averaged across all observations. In the sub-optimizations for $\bm Z^s$ and $\bm \Theta$/$\beta_0$, $\tau$ is treated as a fixed value. Once all other model estimates have been updated, $\tau$ is recalculated with the current model estimates of $\bm Z^s$, $\bm \Theta$, and $\beta_0$ as:
\begin{equation}\label{eq:excess0}
    \widehat{p}_0 = \frac{\sum_{s=1}^S \sum_{i=1}^{n_s} I(y_i^s = 0) - \sum_{s=1}^S \sum_{i=1}^{n_s} \exp [ -\exp (\beta_0 + \bm Z_i^s \bm \Theta)]}{\sum_{s=1}^S n_s}.
\end{equation}

To estimate $\bm B^{{d,s}^{(t)}}$, we first estimate $\bm G^{d^{(t)}}$ for each of the $D$ data views.  For a fixed $\bm \Xi^{{d,s}^{(t-1)}}$, we solve the following optimization problem using using the Adagrad \cite{Adagrad} optimizer in PyTorch \cite{PyTorch}: 
\begin{equation}\label{eq:gopt}
    \widehat{\bm {G}^d}^{(t)} = \min_{\bm G^d \in R^{p_d \times K}}  \sum_{s=1}^S \|(\bm {X}^{d,s} - \bm Z^{s^{(t)}}({{\bm G^d} \cdot {\bm \Xi^{d,s}}}^{(t)})^T\|^2_F + \lambda_G \gamma_d \sum_{l=1}^{p_d} \| {\bm g_l^d}\|_2.
\end{equation}
Convergence is defined as the relative change in \eqref{eq:gopt} evaluated at ${\hat{\bm G}}^{d^{(t)}}$ and ${\hat{\bm G}}^{d^{(t-1)}}$.
We then use these updated estimates for $\bm Z^s$ and $\bm G^d$ to estimate ${\bm \Xi^{d,s}}^{(t)}$ by solving the optimization problem: 
\begin{equation}\label{eq:xiopt}
    {\hat{\bm \Xi}^{{d,s}^{(t+1)}}} = \min_{\bm {\Xi}^{d,s} \in R^{p_d \times K}} \|(\bm {X}^{d,s} - \bm {Z}^{s^{(t)}}(\bm {{G^d}}^{(t)} \cdot {\bm {\Xi}}^{d,s})^T)\|^2_F + \lambda_\xi \gamma_d \sum_{l=1}^{p_d}  \| \bm {\xi}_l^{d,s}\|_2.
\end{equation}
We use the same technique described for the optimization of $\bm G^d$ with an analogous convergence criterion defined as the relative change in \eqref{eq:xiopt} evaluated at $\bm {\hat{\Xi}}^{{d,s}^{(t)}}$ and $\bm {\hat{\Xi}}^{{d,s}^{(t-1)}}$. 

We note that because we use an automatic differentiation algorithm, the $L_{2,1}$ (or block $l_2/l_1$) penalty does not result in zero coefficients. However, the magnitude of the coefficients in $\hat{\bm B}^{d,s}$ for the noise variables are clearly smaller than the coefficients of the signal variables. Thus, we rank and identify relevant variables based on the magnitude of the $L_2$ norm of the rows in $\hat{\bm B}^{d,s}$. In implementing the ranking procedure, the user specifies the number of variables (denote as $N_{top}$) they wish to keep, which could vary across data views. We run Algorithm \ref{ext:alg:zip}  on the full training data and select the  $N_{top}$ variables for each view and subgroup based on the estimated $\bm B^{d,s}$. Then, we run Algorithm \ref{ext:alg:zip} a second time but with the selected  variables. The estimated parameters based on this `subset' of data are used in the prediction procedure described in section \ref{prediction}. 

We estimate $\bm \Theta$ and $\beta_0$ using $\hat{\bm Z^s}^{(t)}$ to optimize the equation problem: 
\begin{equation*}\label{eq:thetaopt}
   ( \hat{\bm \Theta}^{(t)}, \hat{\beta}_0^{(t)}) = \underset{\bm \Theta, \beta_0}{\text{min}} \sum_{s=1}^S F(\bm Y^s, {\bm Z}^{s^{(t)}}, \bm \Theta, \beta_0).
\end{equation*}
We use ISTA (iterative shrinkage-thresholding algorithm) with backtracking \cite{FISTA} to select an appropriate step size. The convergence criterion is the relative change in \eqref{eq:thetaopt} evaluated at $\hat{\bm \Theta}^{(t)}, \hat{\beta}_0^{(t)}$ and $\hat{\bm \Theta}^{(t-1)}, \hat{\beta}_0^{(t-1)}$.

\begin{algorithm}
\footnotesize
\renewcommand{\arraystretch}{0.7}
\caption{Overview of Optimization Algorithm}\label{ext:alg:zip}
\begin{algorithmic}
\State Initialize ${\bm Z^s}^{(0)}$, ${\bm \theta}^{(0)}$, ${\beta_0}^{(0)}$, ${\bm G^d}^{(0)}$, and ${\bm \Xi^{d,s}}^{(0)}$
\If{family = 'ZIP'}
    \State Initialize $\tau^{(0)} = \frac{\sum_{s=1}^S \sum_{i=1}^{n_s} I(y_i^s = 0) - \sum_{s=1}^S \sum_{i=1}^{n_s} \exp [ -\exp ({\beta_0}^{(0)} + {\bm Z_i^s}^{(0)} {\bm \Theta}^{(0)})]}{\sum_{s=1}^S n_s}$
\EndIf

\For{$t = 1, ..., iter_{max}$}
    \For{$s = 1, ..., S$}
        \State \(\displaystyle {\bm Z^s}^{(t)} \gets \argmin_{\bm Z^s} \Big[ F(\bm Y^s, {\bm Z^s}^{(t-1)}, {\beta_0}^{(t-1)}, {\bm \Theta}^{(t-1)}, \tau^{(t-1)}) + 
        \sum_{d=1}^D F(\bm X^{d,s}, {\bm Z^s}^{(t-1)}, {\bm G^d}^{(t-1)}, {\bm \Xi^{d,s}}^{(t-1)}) \Big] \)
        \State Standardize columns of ${\bm Z^s}^{(t)}$ to have mean 0 and variance 1
    \EndFor 

    \For{$d = 1, ..., D$}
        \State \(\displaystyle {\bm G^d}^{(t)} \gets \argmin_{\bm G^d} \sum_{s=1}^S F(\bm X^{d,s}, {\bm Z^s}^{(t)}, {\bm G^d}^{(t-1)}, {\bm \Xi^{d,s}}^{(t-1)}) + \lambda_g \sum_{l=1}^{p_d} \| {g_l^d}\|_2 \)

        \For{$s = 1, ..., S$}
            \State  \(\displaystyle {\bm \Xi^{d,s}}^{(t)} \gets \argmin_{\bm \Xi^{d,s}}  F(\bm X^{d,s}, {\bm Z^s}^{(t)}, {\bm G^d}^{(t)}, {\bm \Xi^{d,s}}^{(t-1)}) + \lambda_\xi  \sum_{l=1}^{p_d} \| {\xi_l^{d,s}}\|_2 \)
        \EndFor 
    \EndFor 

    \State \(\displaystyle \bm \Theta^{(t)}, \beta_0^{(t)} \gets \argmin_{\bm \Theta, \beta_0} \sum_{s=1}^S F(\bm Y^s, {\bm Z^s}^{(t)}, {\beta_0}^{(t-1)}, {\bm \Theta}^{(t-1)}, \tau^{(t-1)}) \) 

    \If{family = 'ZIP'}
        \State $\displaystyle \tau^{(t)} = \frac{\sum_{s=1}^S \sum_{i=1}^{n_s} I(y_i^s = 0) - \sum_{s=1}^S \sum_{i=1}^{n_s} \exp [ -\exp ({\beta_0}^{(t)} + {\bm Z_i^s}^{(t)} {\bm \Theta}^{(t)})]}{\sum_{s=1}^S n_s}$
    \EndIf
    
    \If{Relative Loss $< \epsilon$}
        \State Declare convergence and return estimates
    \ElsIf{$t = iter_{max}$}
        \State Return estimate with warning
    \EndIf
\EndFor 
\end{algorithmic}
\end{algorithm}

\subsection{Tuning Parameters}\label{ext:tuning}

The optimization problem depends on $\lambda=(\lambda_G, \lambda_{\xi})$ and $K$. We use grid and random \cite{bergstra2012random} searches for selecting $\lambda$, and we follow the automatic or scree plot approaches to selecting $K$ described in Butts et al. \cite{butts_2023_HIP}. 
However, we include eBIC (extended Bayesian Information Criterion) \citep{chen_2008_ebic} as an additional model selection criterion. This criterion modifies the priors used in the traditional BIC (Bayesian Information Criterion) to prevent larger probabilities being assigned to models with more covariates. If we consider the set of models with $q$ covariates $\mathcal{M}_q$, then for the $j^{th}$ model $\mathcal{M}_{qj} \in \mathcal{M}_q$ with maximum likelihood estimated parameters $\widehat{\theta}(\mathcal{M}_{qj})$, eBIC is defined as 
$$eBIC_\delta(\mathcal{M}_{qj}) = - 2 \log \mathcal{L}_n({\widehat{\theta}(\mathcal{M}_{qj}))} + \nu(\mathcal{M}_{qj}) \log (n) + 2 \delta \log (\kappa (\mathcal{M}_q))$$ 
for $0 \leq \delta \leq 1$ where $\nu(\mathcal{M}_{qj})$ is the number of estimated parameters, $\kappa (\mathcal{M}_q)$ is the number of models in $\mathcal{M}_q$, and $n$ is the number of observations. When $\delta = 0$, eBIC is equivalent to the standard BIC, and $\delta = 1$ ensures consistency when the number of covariates is very large.

The eBIC criterion for HIP is defined in \eqref{eq:eBIC}. We use $F(\bm Y^s, \widehat{\bm Z}^s, \widehat{\bm \Theta}, \widehat{\beta_0})$ in the first term as it is proportional to the log-likelihood. For the number of parameters $\nu$, we count the number of variables that will be included in the subset model fit for HIP. A variable is included in the subset fit if it is one of the $N_{top}$ variables selected in at least one subgroup, where $N_{top}$ is user-specified (e.g., top 10\% of variables or top 50 variables) and can vary by data view. Thus, we define $\nu(\widehat{\bm B}) = \sum_{d=1}^D \sum_{l=1}^{p_d} I(\widehat{\bm B}_l^{d,s} \in N_{top} \text{ for any }s = 1, ..., S)$; note this value is the same for all subgroups. For the number of models in $\mathcal{M}_q$, we note that in a given view, we could include as few as $N_{top}$ variables in the case where all subgroups select the same set of variables and as many as $S*N_{top}$ variables when all subgroups select distinct sets of variables. Thus, to count the size of $\mathcal{M}_q$ where $q = N_{top}$, we sum the combinations of choosing each possibility between $N_{top}$ and $S*N_{top}$ from the $p_d$ variables in the view. The code will return values for $eBIC_0$, $eBIC_{0.5}$, and $eBIC_1$:

\begin{equation}\label{eq:eBIC}
    eBIC_\delta = 2 \sum_{s=1}^S F(\bm Y^s, \widehat{\bm Z}^s, \widehat{\bm \Theta}, \widehat{\beta_0}) + \sum_{s=1}^S \log(n_s) \nu(\widehat{\bm B}) + 2 \delta \sum_{d=1}^D \log \Big[ \sum_{w = N_{top}}^{S*N_{top}} {p_d \choose w} \Big]
\end{equation}
\section{Simulations}\label{ext:sec:sims}

\subsection{Set-up}
Simulations were run for binary, Poisson, and ZIP outcomes. For all outcomes, there were two views and two subgroups, i.e., $D = S = 2$, and there were $n_1 = 250$ subjects belonging to the first subgroup and $n_2 = 260$ subjects belonging to the second subgroup. We considered two different scenarios for degree of subgroup heterogeneity: Full Overlap and Partial Overlap. In both scenarios, there are 50 variables important to each subgroup for each view. In the Full Overlap scenario, the 50 important variables are the same for both subgroups; in the Partial Overlap scenario, 25 of the important variables are common to both subgroups and 25 are unique to each subgroup. For each of the outcomes and scenarios, there were two different sets of variable dimensions in the data views. In the low dimensional setting, the first data set had $p_1 = 300$ variables, and the second had $p_2 = 350$. In the high dimensional setting, the first data set had $p_1 = 2000$ variables, and the second had $p_2 = 3000$ variables. 

\subsection{Data Generation}
The data were generated following a process based on Luo et al. \citep{luo2016canonical}. Entries in rows of $\bm {B}^{d,s}$ corresponding to a signal variable were drawn from a $U(-1, -0.5) \cup U(0.5,1)$; otherwise the entry was set to 0. The columns of each $\bm {B}^{d,s}$ were then orthonormalized using a QR decomposition. The entries of $\bm Z^s$ are drawn from $N(\mu = 25.0, \sigma = 3.0)$ and entries of $\bm {E}^{d,s}$ from $N(\mu = 0.0, \sigma = 1.0)$. Then the covariate matrices $\bm X^{d,s}$ are formed as $\bm Z^s \bm B^{{d,s}^T} + \bm E^{d,s}$. 

At this point, the processes diverge somewhat for the different outcomes. For the binary outcome, we apply the \textit{softmax} function to $\bm W^s = \mathcal{J}_{n_s} \beta_0  + \bm Z^s \bm \Theta + \bm E_y^s$ where $\mathcal{J}_{n_s}$ is an $n_s \times 1$ matrix of ones, $E_y^s$ is an $n_s \times 2$ matrix of standard normal errors and assign the class with the largest probability for each observation. Here $\beta_0 = \begin{bmatrix} 0.5 & 0.5 \end{bmatrix}$ and $\bm \Theta = \begin{bmatrix} 1.0 & 0.5 \\ 0.2 & 0.8 \end{bmatrix}$. For the Poisson and ZIP outcomes, we first standardize the columns of $\bm Z^s$ to have mean $0$ and variance $1$. The observations are then generated from the Pytorch \citep{PyTorch} Poisson random variable generator with mean $\exp(\mathcal{J}_{n_s}\beta_0 + \bm Z^s \bm \Theta)$, i.e., observation $y_i^s \sim Poisson(\exp(\beta_0 + \bm Z_i^s \bm \Theta))$. Here $\beta_0 = 2.0$ and $\bm \Theta = \begin{bmatrix} 0.7 & 0.2 \end{bmatrix}^T$. For the ZIP outcome, each observation is then multiplied by a draw from a Bernoulli distribution that is 0 with probability $\tau = 0.25$.

\subsection{Comparison Methods}\label{ext:sec:sims:compare}
For all outcomes under consideration (binary, Poisson, and ZIP), we are unaware of any other methods that perform integrative analysis while also accounting for subgroup heterogeneity. For the binary outcome, we compare HIP to two integrative analysis methods: CVR \citep{luo2016canonical} as implemented in the R package \emph{CVR} \citep{cvrpackage} and SIDA \citep{SIDA} as implemented in the R package \emph{mvlearnR} \citep{mvlearnr}. Because neither of these methods accounts for subgroup heterogeneity, we implement each method two ways: (1) all subgroups concatenated within each view (Concatenated) and (2) a separate model for each subgroup (Subgroup). We also compare HIP to the Lasso \citep{lasso} and the Elastic Net \citep{enet} as implemented in the R package \emph{glmnet} \citep{glmnet}. Neither of these two methods perform integrative analysis, so the two views are concatenated for these methods. We again implement two ways: (1) concatenating the subgroups (Concatenated) and (2) separate models for each subgroup (Subgroup).

For the Poisson outcome, SIDA is no longer applicable, so we instead compare HIP to the two-step integrative analysis method SELPCCA \citep{safo_2018_selpcca} as implemented in the R package \emph{mvlearnR} \citep{mvlearnr}. Since SELPCCA does not account for subgroup heterogeneity, we again fit concatenated and subgroup models (Concatenated SELPCCA and Subgroup SELPCCA, respectively). 

For the ZIP outcome, CVR, Lasso, and Elastic Net cannot explicitly account for a ZIP outcome, but we still fit these models using a Poisson family. We also fit HIP specifying a Poisson outcome [HIP (Grid)-Poisson and HIP (Random)-Poisson] to demonstrate the importance of accounting for the zero-inflated nature of the data. Finally, because SELPCCA is a two-step method, we use the canonical variables from SELPCCA in a ZIP regression model fit with the {\fontfamily{qcr}\selectfont zeroinfl} function \citep{zeroinfl} in R package \emph{pscl} \citep{pscl}.  We again fit a model on the concatenated subgroups (Concatenated SELPCCA-ZIP) and separate models for each subgroup (Subgroup SELPCCA-ZIP).

Tuning parameters for the comparison methods were selected using 10-fold cross-validation. For the Elastic Net, we set $\alpha = 0.5$. Tuning parameters for HIP were selected using $eBIC_1$; we considered a range of $(0,2]$ for $\lambda_\xi$ and $\lambda_G$ with $8$ steps for each. The true values for $K$ (i.e., $2$) and $N_{top}$ (i.e., $50$) were used for all simulations.

\subsection{Evaluation Measures}
We compare HIP to the existing methods in terms of variable selection and prediction ability for new data. For variable selection, we will estimate the true positive rate ($\text{TPR} = \frac{\text{True Positives}}{\text{True Positives + False Negatives}}$), false positive rate ($\text{FPR} = \frac{\text{False Positives}}{\text{True Negatives + False Positives}}$), and F$1$ score ($\text{F1} = \frac{\text{True Positives}}{\text{True Positives} + \frac{1}{2}\text{(False Positives + False Negatives)}}$) which are all constrained to be between 0 and 1. Ideally, TPR and F$1$ are $1$ and FPR is $0$.

To compare predictive ability for the binary outcome, we look at classification accuracy, and for both Poisson and ZIP outcomes, we look at the fraction of deviance explained, $D^2 = \frac{D_{null} - D_{opt}}{D_{null}}$, where $D_{null}$ is the deviance of the null model and $D_{opt}$ is the deviance of the model with optimal tuning parameters \citep{hastie2016learning}. Results are averaged over $20$ Monte Carlo data sets.  

\subsection{Results}

We focus on the results for the ZIP outcome here as the motivating data have a ZIP outcome; results for the ZIP outcome with Full Overlap scenario as well as binary and Poisson outcomes are presented in the supplemental information (Figures S1 - S8).

\begin{figure}[htbp]
    \centering
    \caption{{\bf Results for ZIP Outcome, Full Overlap Scenario.} The first row corresponds to the low dimension ($p_1$ = 300, $p_2$ = 350) and the second to the high dimension ($p_1$ = 2000, $p_2$ = 3000). For all settings, $n_1 = 250$ and $n_2 = 260$. The right column is the fraction of deviance explained ($D^2$), so a higher value indicates better performance. Results are mean $\pm$ one standard deviation summarized across 20 generated data sets.}\label{ext:fig_perf_zip_full}
    \includegraphics[width=\textwidth]{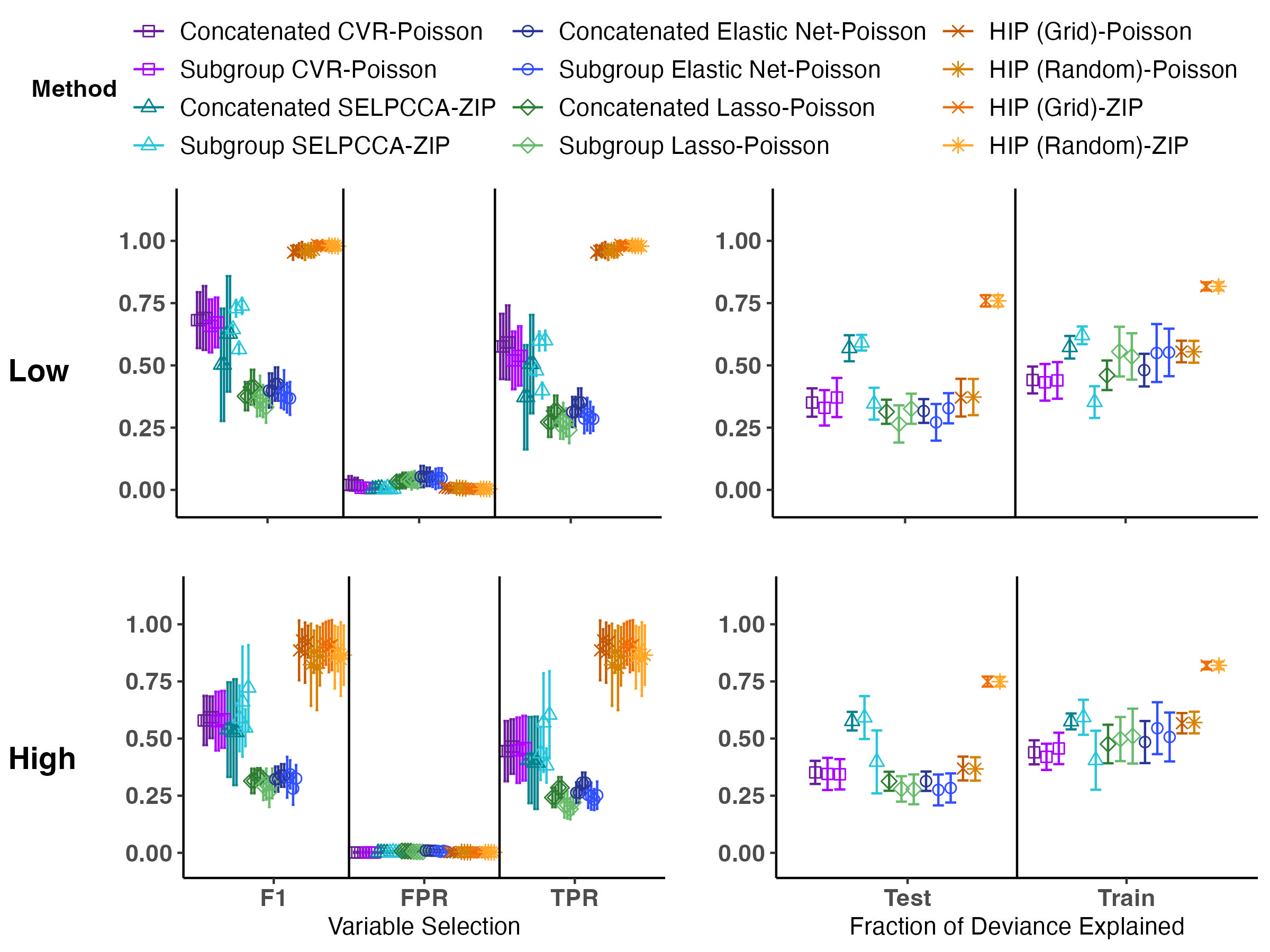}
\end{figure}

\begin{figure}[htbp]
    \centering
    \caption{{\bf Performance Results for ZIP Outcome, Partial Overlap Scenario.} The first row corresponds to the low dimension ($p_1$ = 300, $p_2$ = 350) and the second to the high dimension ($p_1$ = 2000, $p_2$ = 3000). For all settings, $n_1 = 250$ and $n_2 = 260$. The right column is the fraction of deviance explained ($D^2$), so a higher value indicates better performance. Results are mean $\pm$ one standard deviation summarized across 20 generated data sets.}\label{ext:fig_perf_zip_partial}
    \includegraphics[width=\textwidth]{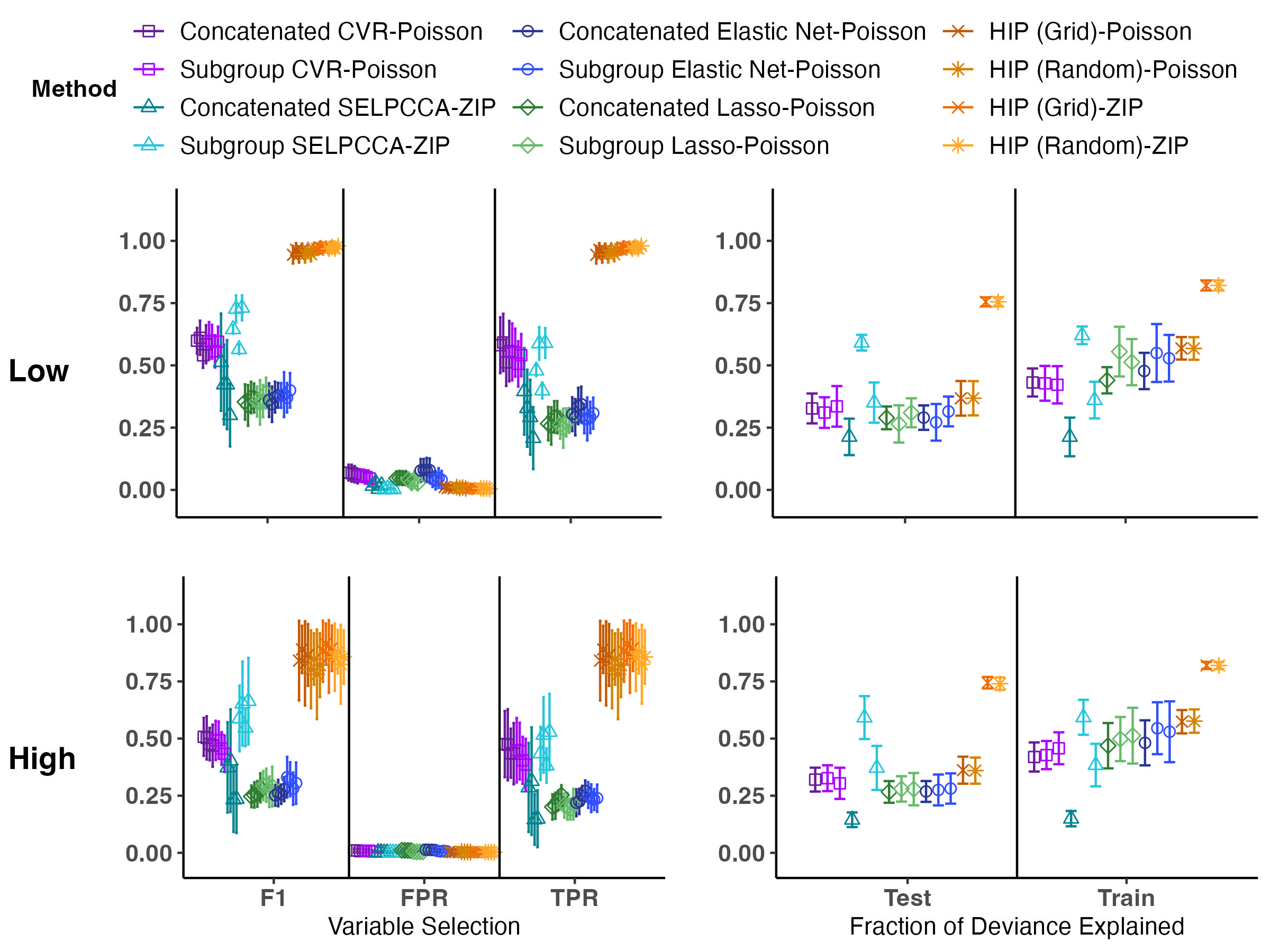}
\end{figure}

First, we note the performance of HIP (Grid) and HIP (Random) are very similar in both the low and high dimensional settings for both the Full and Partial Overlap scenarios (Figures \ref{ext:fig_perf_zip_full} and \ref{ext:fig_perf_zip_partial} respectively), so we recommend the use of HIP (Random) as it is computationally faster. In terms of variable selection, HIP (Grid) and HIP (Random) show the highest TPR and F$1$ values; all methods show low FPRs. The ZIP model only improves variable selection slightly over the HIP Poisson fits. The other two integrative methods, CVR and SELPCCA-ZIP, show similar variable selection performance to each other and show a benefit over the non-integrative methods (Elastic Net and Lasso), but they do not perform nearly as well as HIP. All of the comparison methods seem to be missing many of the true signal variables as evidenced by lower TPRs. 

In terms of predictive ability in the Full Overlap Scenario, we note that HIP and the SELPCCA-ZIP models have a much higher fraction of deviance explained than the methods that do not account for the zero-inflated nature of the data, highlighting the importance of doing so. In the Partial Overlap Scenario, Concatenated SELPCCA-ZIP has a much lower $D^2$ than in the Full Overlap Scenario suggesting that this method has more difficulty when subgroup heterogeneity exists in the data. HIP has the highest fraction of deviance explained out of all methods applied for both the low and high dimensional settings in both the Full and Partial Overlap Scenarios indicating HIP is still favorable even if subgroup heterogeneity does not exist in the data.

Computation times for all methods are summarized for the Full and Partial Overlap Scenarios in Supplemental Figures S9 and S10 respectively. The Lasso and Elastic net are the fastest in all scenarios, but because these methods do not perform integrative analysis or account for subgroup heterogeneity, the variable selection and predictive ability suffers. Of the methods that perform integrative analysis, HIP (Random) is consistently the fastest with larger computational advantages in the high dimensional setting particularly when compared to CVR. Concatenated SELPCCA-ZIP has similar computation times as HIP (Random) in the Partial Overlap setting, but this is the setting where the performance of Concatenated SELPCCA-ZIP was reduced dramatically.

\section{Application to Exacerbation Frequency}\label{ext:sec:rda}

\subsection{Goals}
In this section, our goal is to use the genetic and proteomic data from the COPDGene Study \citep{regan2011genetic} in combination with clinical data to gain new insights into the molecular architecture of COPD in males and females. To be included in the analyses, participants had to have COPD at P2 (defined as GOLD stage $\ge 1$) and have proteomic, genomic, and selected clinical covariates (age, BMI, race, pack-years, FEV$_1$\%, AWT, and \% emphysema) available. There were $N = 1374$ participants meeting these criteria with $n_1 = 780$ males and $n_2 = 594$ females; demographic characteristics of this sample are in Table \ref{ext:tab:rda:dem}. Continuous variables were compared between males and females using t-tests and categorical variables using $\chi^2$ tests. Participants were predominantly non-Hispanic white; there were not statistically significant sex differences in age, BMI, BODE index, percentage of current smokers, or percentage with diabetes. Males and females differed in their exacerbation frequency (p $<0.001$) but did not have significantly different lung function as measured by mean FEV$_1$\% predicted. Given the available data, and the sex differences in exacerbation frequency, we first  identify genes and proteins common and specific to males and females associated with exacerbation frequency. Using those identified genes and proteins, we then explore pathways enriched for males and females.

\input{Tables/Tab_rda_dem}

\subsection{Applying HIP and Existing Methods}
The COPDGene data includes 19263 genes and 4979 proteins. We first performed unsupervised filtering to select the 5000 genes and 2000 proteins with the largest variances. To preserve as much generalizability as possible, we then randomly split the data 50 times into train (75\%) and test (25\%) data sets keeping the proportions of males and females the same. Genes and proteins that were consistently selected across these splits were considered ``stable" as these would be most likely to show consistent findings. Within each split, we performed supervised filtering using the training data by regressing exacerbation frequency on each of the genes and proteins retained after the unsupervised filtering. We used the {\fontfamily{qcr}\selectfont zeroinfl} function \citep{zeroinfl} where the current gene or protein was the only predictor in the count model and only an intercept for the zero model. Genes and proteins with an uncorrected p-value $< 0.10$ were included in the models. Because this supervised filtering was repeated for each split, a different set of variables could enter the models for each split of the data. 

For existing methods, we fit the subgroup implementations described in Section \ref{ext:sec:sims:compare} for CVR (using a Poisson family), SELPCCA-ZIP, Lasso (using a Poisson family), and Elastic Net (using a Poisson family). For HIP, we applied HIP (Grid) and HIP (Random) using the ZIP family and using the Poisson family [HIP (Grid)-Poisson and HIP (Random)-Poisson]. Additionally, we fit HIP using the ZIP family with an additional clinical data view that was not penalized [HIP (Grid)-ZIP+Clinical and HIP (Random)-ZIP+Clinical]. We used the training data to select tuning parameters and calculate training $D^2$ and then used the test data to calculate a test $D^2$.

We defined the ``stable" genes and proteins by ranking them based on the product of (a) the number of splits in which the variable was included in the $N_{top}$ variables and (b) the number of splits in which the variable was included in the $N_{top}$ variables divided by the number of splits in which the variable was entered into the models. The top 1\% of genes and proteins for males and females based on this ranking were identified as the ``stable" genes and proteins for each method.

To select tuning parameters, CVR, SELPCCA, Lasso, and Elastic Net used 10-fold cross-validation. For HIP, when we used the $\lambda$ range $(0,2]$ used in simulations, the upper bound was consistently being selected, so we increased the range until this was not happening resulting in a range of $(0,15]$ for $\lambda_G$ and $\lambda_\xi$. The best model was selected using $eBIC_1$. We also needed to specify $K$, the number of latent components for HIP and an equivalent parameter for CVR. The automatic approach specified in Butts et al.  \cite{butts_2023_HIP} with a threshold of $0.20$ on the concatenated data suggested $K=3$. On the separate $\bm X^{d,s}$, it suggested $K=3$ for the genes and $K=1$ for the proteins. Scree plots for the concatenated and separate data are in Supplemental Figure S11. We thus selected $K=3$ components for HIP and CVR. For HIP, we set $N_{top} = 125$ genes and $50$ proteins.

\subsection{Results}

\subsubsection{Fraction of Deviance Explained and Computation Times}

\begin{figure}[htbp]
    \centering
    \caption{{\bf Fraction of Deviance Explained Across 50 Data Splits of the COPDGene Data.} Violin plots display the distribution of fraction of deviance explained across the 50 data splits. For each of CVR, SELPCCA-ZIP, Lasso, and Elastic Net, there is a violin plot for each subgroup as comparator methods require separate models for each subgroup to allow for possible subgroup heterogeneity.}\label{ext:fig_rda_dev}
    \includegraphics[width=\textwidth]{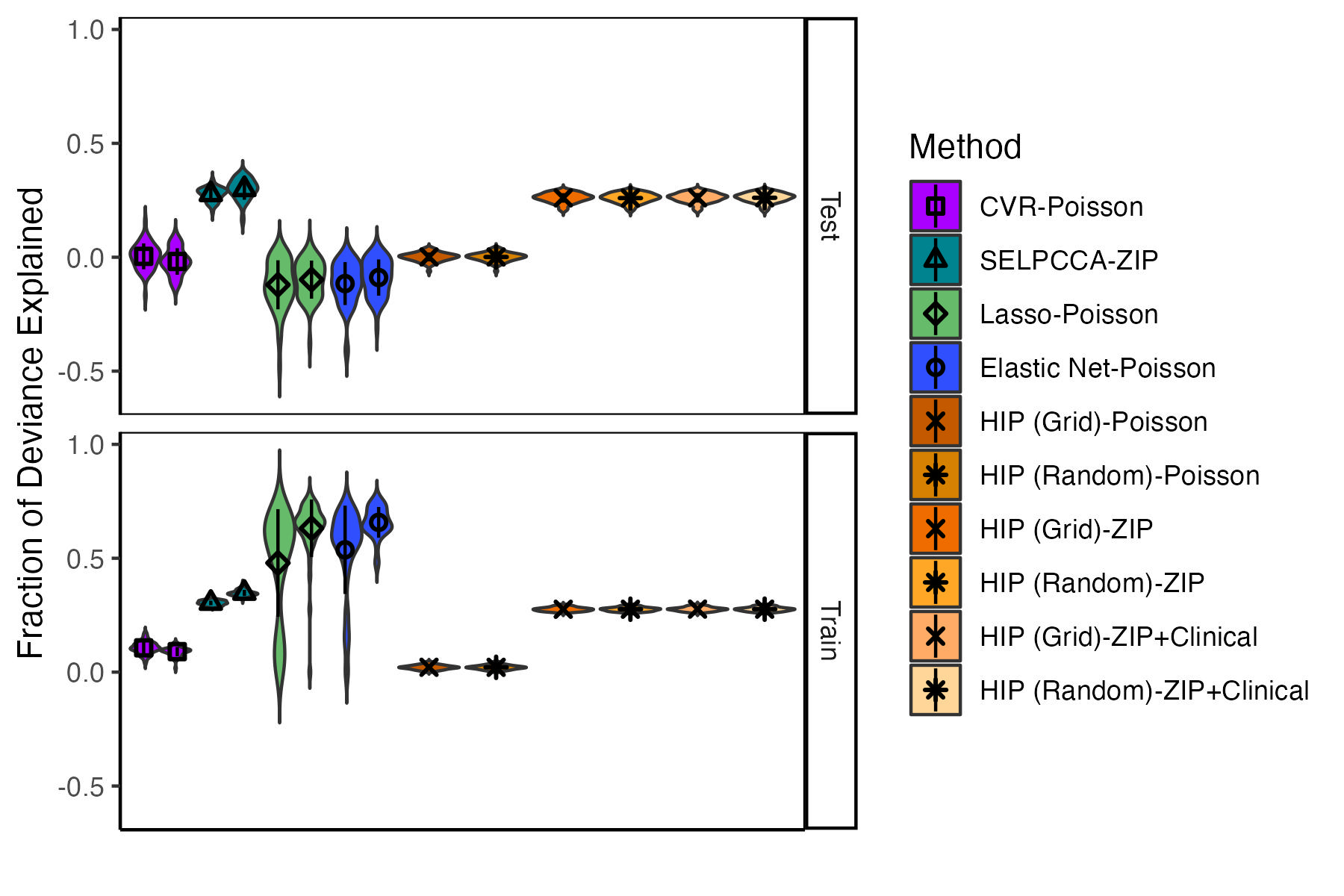}
\end{figure}

Figure \ref{ext:fig_rda_dev} shows violin plots of the train and test fraction of deviance explained for the 50 Train/Test data splits. As expected, HIP (using ZIP model) and SELPCCA-ZIP, the two methods that account for the zero-inflated nature of the data, have the best predictive ability. The number of variables selected in the 50 Train/Test splits are summarized in Supplemental Table S1. Similarly, Supplemental Figure S12 shows a violin plot of the run times for each method. Of the methods performing integrative analysis, HIP (Random) regardless of family tended to have the fastest computation times, although SELPCCA-ZIP often had similar run times.

\subsubsection{Selected Genes and Proteins}

Supplemental Table S2 shows the number of ``stable" genes and proteins common and specific to males and females identified by each method. Supplemental Figure S13 shows the overlap of ``stable" genes and proteins selected for males and females by each method. There are few overlaps in the ``stable" genes and proteins, but the overlaps that do occur tend to be in the methods that perform integrative analysis (i.e., HIP, CVR, and SELPCCA) suggesting integrative methods may result in more reproducible findings.

HIP (Random) and HIP (Grid) showed strong agreement in the ``stable" genes and proteins (Supplemental Table S3) which also supports the use of the random search instead of the grid search. The ``stable" genes selected by HIP (Random) for males and females and their average estimated weights are in Supplementary Tables S4 and S5 respectively. Analogous results for proteins are presented in Supplementary Table S6. The weights in these tables are averages of the estimated weights ($L_2$ norm of rows in $\widehat{\bm B}^{d,s}$) from the subset fit in the splits where the variable was included in the subset fit (i.e., the variable was in $N_{top}$ for at least one subgroup).

The top gene for males was activating signal cointegrator 1 complex subunit (ASCC2). Wilson et al. \cite{wilson_2020_ASCC2} looked  for genes that were differentially expressed in COPD patients with and without cachexia (a loss of weight and muscle) and identified ASCC2 in a sample of 400 COPDGene Study participants and replicated the finding in a sample of 114 participants from the ECLIPSE Study; cachexia occurs more frequently in those with more advanced COPD. The top gene for females was dematin actin binding protein (DMTN). Lee et al. \cite{lee_2016_DMTN} measured DNA methylation on 100 participants (60 with and 40 without COPD) from a Korean COPD Cohort and identified DMTN (also known as EPB49) as a differentially methylated region when comparing current smokers to never smokers; the authors also note this gene has been identified in previous epigenome-wide association studies of smoking. Thus this gene may be a candidate for future research to investigate the relationship with COPD specifically.

The top protein for males was SHC adaptor protein 1 (SHC1) (ranked fifth for females). Li et al. \cite{li_2021_SHC1} ranked candidate genes based on gene risk scores where a higher rank indicated a stronger relationship to COPD. The SHC1 gene was ranked third out of 200 candidate genes and had lower expression levels in patients with COPD compared to healthy controls. The top protein for females was amyloid beta precursor protein (APP) (ranked second for males). Almansa et al. \cite{almansa_2012_APP} compared gene expression levels of 12 patients with COPD requiring treatment in the ICU compared to 16 patients with COPD who were admitted to the hospital for treatment but did not require the ICU. They found that the patients admitted to the ICU showed higher expression levels of APP compared to patients who were not admitted to the ICU.

\subsubsection{Pathway Analysis}

We tested for overrepresentation of pathways in our ``stable'' proteins and genes for males and females using Ingenuity Pathway Analysis \citep{IPA}. Table \ref{ext:tab:rda:pathways} shows the top 10 canonical pathways for genes and proteins for males and females.

There were both common and subgroup-specific gene pathways for males and females. The top gene pathway for males was the CLEAR Signaling Pathway, and the top gene pathway for females was the STAT3 pathway.  The STAT3 pathway is known to be involved in inflammatory responses to many diseases including COPD \citep{kiszalkiewicz_2021_STAT3}. The Iron homeostasis signaling pathway that was found in the AWT analysis in Butts et al. \cite{butts_2023_HIP} for both males and females was again present in the top gene pathways for males and females. 

There was complete overlap in the top 10 protein pathways for males and females; this makes some sense because of the 20 stable proteins identified for males and females, 17 of them overlapped. For both males and females, the top pathway was the wound healing signaling pathway followed by PDGF signaling. The PDGF family includes PDGFA (platelet derived growth factor subunit A) and PDGFB (platelet derived growth factor subunit B) and is associated with wound healing; when PDGFs are found outside the context of wound healing, it seems to contribute to many diseases \citep{Kardas_2020_PDGF}. Though these authors focus on asthma rather than COPD, they also state that PDGF is expressed in airway epithelial cells and PDGFB is expressed in inflamed airway tissue.

\input{Tables/Tab_rda_pathways}

\clearpage
\section{R Shiny Application}\label{sec:shiny}

\begin{figure}
    \caption{{\bf HIP Shiny Application Screens}}
    \begin{subfigure}
        \centering
        \caption{`About' Tab}\label{fig:shiny:about}
        \includegraphics[width=\textwidth]{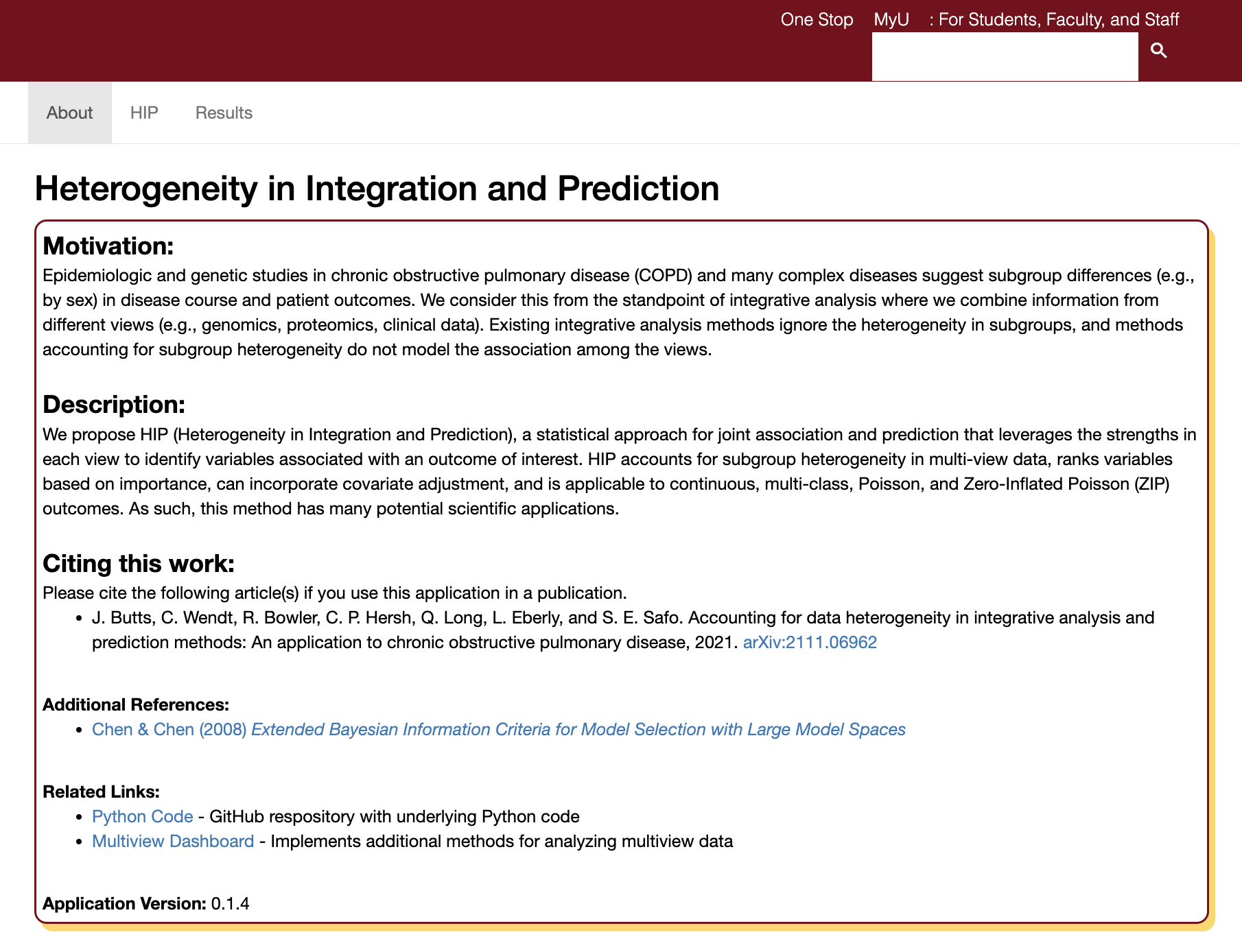}
    \end{subfigure} \\
    \begin{subfigure}
        \centering
        \caption{Input Data - COVID-19 Example}\label{fig:shiny:upload}
        \includegraphics[width=\textwidth]{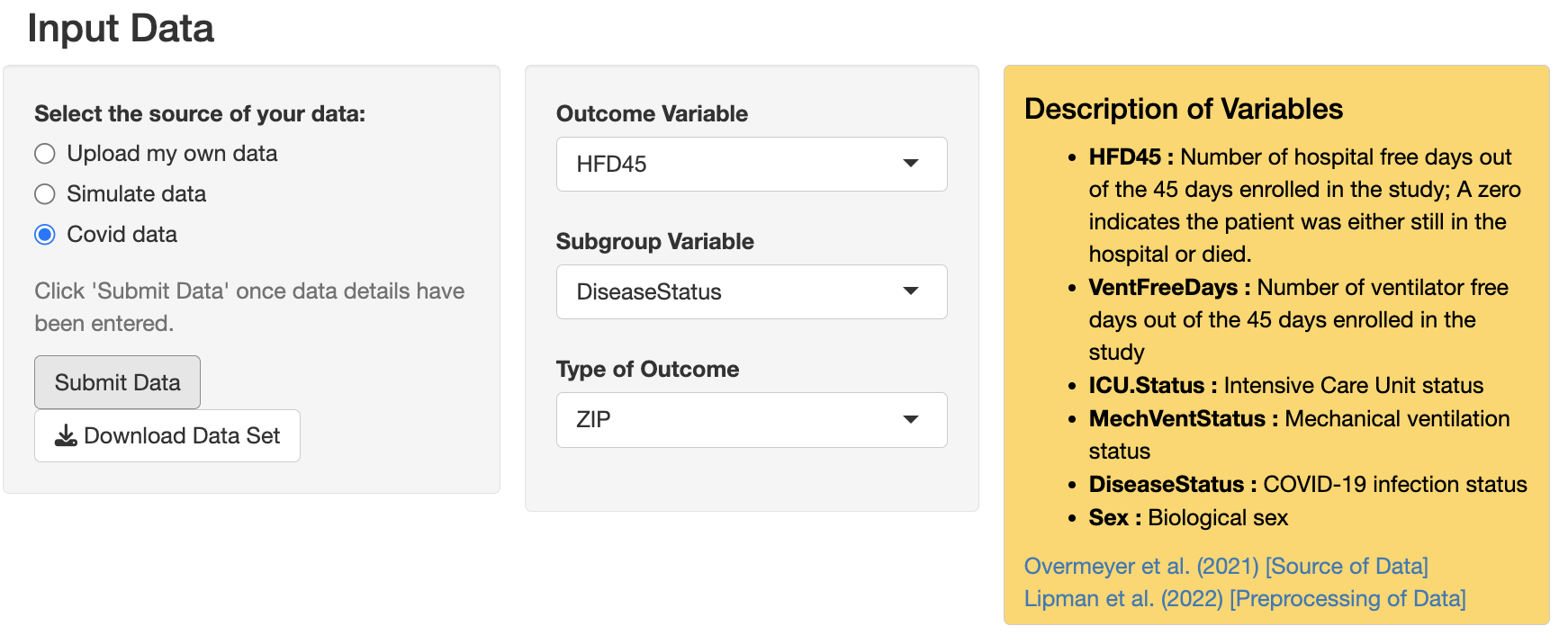}
    \end{subfigure}
\end{figure}

While HIP and its extension to additional outcomes has many research areas to which it could be applied (within and outside of human health), it is implemented in Python which may be a barrier to some researchers without a coding background. We introduce an R Shiny \citep{r_shiny} application that provides a graphical interface to apply HIP to data that is either uploaded, simulated, or available within the application. In this section we provide a brief overview and highlights of the application. A detailed example of using the application is in the supplemental information. The application is accessible on shinyapps.io at \url{https://multi-viewlearn.shinyapps.io/HIP_ShinyApp/}.

The Shiny application has three tabs. The first tab, `About', provides a brief overview of the method and related links (Figure \ref{fig:shiny:about}). The second tab, `HIP', is where the user will set-up the data and parameters for the method. We provide options for the user to upload their own data, simulate data (based on our simulation examples) and implement HIP on an example COVID-19 data (Figure \ref{fig:shiny:upload}). The third tab, `Results', is where the user will submit the analysis and see results. This tab produces outputs of the model including prediction information and variable importance to help the user understand the results. 

Once the `Run Analysis' button has been clicked, a progress notification (Supplementary Figure S19) will appear in the lower right corner of the screen so the user knows the analysis has started. The user will know the analysis is complete when the progress notification disappears and the run time is displayed to the right of the `Run Analysis' button. 
After the analysis is complete, the left column in the `Result Summary' section will provide a button to download results, and the right column will display some basic information about the results for the user including the convergence status, the $\lambda$ values used, the $eBIC_0$ or other selection criterion value, the run time, and the applicable training prediction metric (Supplementary Figure S20). 
We also show prediction metric for the training data and the test data if they are available. If there are no test data available, a message will be printed for the user stating so. The prediction metric is the mean squared error for Gaussian outcomes, classification accuracy for multi-class outcomes, and fraction of deviance explained for Poisson and ZIP outcomes. The appropriate metric and formula are displayed in a gold box in the left column for users.

The weights from the estimated $\bm B^{d,s}$ matrices are displayed in two formats. The first is a variable importance plot that includes all variables that were included in the subset model fit (Supplementary Figure S22). The variables are ranked based on the weights within each view and subgroup.
The second output is an interactive table with columns for the rank (within view and subgroup), variable, weight, view, and subgroup (Supplementary Figure S23). This table can be sorted by or filtered on any column so that the user can explore the results. Some possible options that could be of interest include selecting the top ranked variable(s) for each view and subgroup, selecting a specific variable to compare between subgroups, or selecting all results for a single view and subgroup (Supplementary Figures S24-S26). 

In general, user inputs are in gray boxes, generated outputs have a white background, and information for the user is in gold boxes. Please refer to the supplementary material for detailed descriptions of HIP and an illustration of HIP on a publicly available data on COVID-19.

\clearpage
\section{Conclusion}\label{ext:sec:conc}

In this paper, we have extended HIP as proposed in Butts et al. \cite{butts_2023_HIP} to accommodate multi-class, Poisson, and zero-inflated Poisson (ZIP) outcomes allowing researchers to investigate additional clinically relevant outcomes in an integrative analysis framework that also accounts for subgroup heterogeneity. We retain the benefits of a joint association and prediction method for data integration to select clinically meaningful subgroup-specific and common features and the ability to include clinical covariates. In simulations, HIP demonstrated improved variable selection abilities for binary, Poisson, and ZIP outcomes compared to existing methods. While all methods showed similar classification accuracy in the binary outcome simulations, HIP showed small improvements in $D^2$ in the Poisson outcome simulations and substantial improvements in $D^2$ in the ZIP outcome simulations. When applied to data from the COPDGene Study, we were able to identify common and subgroup-specific genes and proteins associated with exacerbation frequency; previous literature has identified at least some of these as being related to COPD.

The R Shiny Application developed here provides a user-friendly interface to apply HIP to any data with multiple data views, measured on pre-specified subgroups, being used to predict an outcome. In the first tab, the application introduces HIP and describes why HIP may be a desirable method to consider for analyzing a multi-view data set with potential subgroup heterogeneity. In the second tab, the application guides the user through inputting data, selecting an appropriate $K$, and setting parameters for running HIP. Finally, the third tab produces outputs of the model including prediction information and variable importance to help the user understand the results.

There are still some limitations to HIP requiring further research. As in the originally proposed HIP \citep{butts_2023_HIP}, $N_{top}$, the number of variables to keep for each view, must be specified. In simulations, we know the true value of $N_{top}$, but we cannot know the true value in applications to real data. Sensitivity analyses could help investigate the performance when values other than the true value are specified. Additionally, the inclusion of clinical covariates in the COPDGene application did not improve predictive abilities on the test data. The reason for this is unclear and could be for multiple reasons. One possibility is the variables in the clinical view do not explain additional variation in the outcome beyond the genes and proteins. Another possibility is that the large difference in the number of variables in this view compared to the other views affects the estimation. Despite these limitations, this extension to HIP allows researchers to explore a wide variety of new questions by considering multi-class, Poisson, or ZIP outcomes that are clinically meaningful.


\section*{Supporting information}

Additional supporting information may be found in the online version of the article at the publisher’s website. The  Python code for implementing HIP is publicly available on GitHub at \url{https://github.com/lasandrall/HIP}. An R-package for HIP would also be made available at this same link. 

\section*{Acknowledgements}
This work was supported NIGMS grant 1R35GM142695, NHLBI grants U01 HL089897 and U01 HL089856 and by NIH contract 75N92023D00011. The COPDGene study (NCT00608764) is also supported by the COPD Foundation through contributions made to an Industry Advisory Committee that has included AstraZeneca, Bayer Pharmaceuticals, Boehringer-Ingelheim, Genentech, GlaxoSmithKline, Novartis, Pfizer, and Sunovion.
\bibliographystyle{unsrtnat}
\bibliography{02_References}

\end{document}

%% file: Tables/Tab_rda_dem.tex

\begin{table}
\centering
\renewcommand{\arraystretch}{0.7}
\setlength{\tabcolsep}{12pt}
\caption{{\bf Characteristics of Participants Included in COPDGene Application.} All measurements are from the Year 5 study visit to align with the collection of proteomic and genomic data.}\label{ext:tab:rda:dem}
\begin{tabular}{lccc}
\toprule
\multicolumn{1}{c}{Variable} & \multicolumn{1}{c}{Males} & \multicolumn{1}{c}{Females} & \multicolumn{1}{c}{P-value} \\
  & N = 780 & N = 594 &  \\
\midrule
Age & 68.29 (8.35) & 68.03 (8.36) & 0.564\\
BMI & 28.03 (5.62) & 27.69 (6.59) & 0.318\\
FEV$_1$ \% Predicted & 62.03 (22.94) & 62.91 (22.59) & 0.474\\
BODE Index & 2.43 (2.44) & 2.63 (2.38) & 0.151\\
\% Emphysema & 11.27 (11.84) & 9.39 (11.45) & 0.003\\
Pack Years & 53.05 (26.63) & 47.57 (24.99) & $<$0.001\\
Airway Wall Thickness & 1.17 (0.23) & 1.00 (0.21) & $<$0.001\\
Exacerbation Frequency & 0.37 (0.91) & 0.54 (1.00) & 0.002\\
Non-Hispanic White (\%) & 82 & 78 & 0.089\\
Current Smoker (\%) & 66 & 65 & 0.532\\
Diabetes (\%) & 17 & 14 & 0.197\\
\bottomrule
\multicolumn{4}{l}{\footnotesize COPD = Chronic Obstructive Pulmonary Disease}\\
\multicolumn{4}{l}{\footnotesize BMI = Body Mass Index}\\
\multicolumn{4}{l}{\footnotesize FEV$_1$ = Forced Expiratory Volume in 1 Second}\\
\multicolumn{4}{l}{\footnotesize BODE = \underline{B}ody mass index, airflow \underline{O}bstruction, \underline{D}yspnea, and \underline{E}xercise capacity}\\
\end{tabular}
\end{table}

%% file: Tables/Tab_rda_pathways.tex

\begin{sidewaystable}
    \centering
    \scriptsize
    \renewcommand{\arraystretch}{0.8}
    \caption{{\bf Top 10 Canonical Pathways.}}\label{ext:tab:rda:pathways}
\begin{tabular}{ccllc}
\toprule
View	&	Subgroup	&	Canonical Pathway	&	Molecules	&	Unadjusted P-value	\\
\midrule
\multirow{20}{*}{\centering Genes}	&	\multirow{10}{*}{\centering Males}	&	CLEAR Signaling Pathway	&	ATP6V0C,IGF2R,PINK1,PPP2R5B	&	0.001	\\
	&		&	STAT3 Pathway	&	IGF2R,IL1R2,PIM1	&	0.002	\\
	&		&	IL-10 Signaling	&	BLVRB,IL1R2,PBX1	&	0.002	\\
	&		&	Methylglyoxal Degradation I	&	HAGH	&	0.005	\\
	&		&	Heme Degradation	&	BLVRB	&	0.007	\\
	&		&	Histidine Degradation III	&	HAL	&	0.013	\\
	&		&	$\gamma$-glutamyl Cycle	&	ANPEP	&	0.020	\\
	&		&	Iron homeostasis signaling pathway	&	ATP6V0C,SLC25A37	&	0.023	\\
	&		&	Histidine Degradation VI	&	HAL	&	0.039	\\
	&		&	Granulocyte Adhesion and Diapedesis	&	C5AR1,IL1R2	&	0.042	\\

\cmidrule{2-5}
         
	&	\multirow{10}{*}{\centering Females}	&	STAT3 Pathway	&	IGF2R,IL1R1,PIM1	&	0.001	\\
	&		&	Iron homeostasis signaling pathway	&	ALAS2,ATP6V0C,SLC25A37	&	0.001	\\
	&		&	Methylglyoxal Degradation I	&	HAGH	&	0.005	\\
	&		&	Heme Degradation	&	BLVRB	&	0.006	\\
	&		&	Tetrapyrrole Biosynthesis II	&	ALAS2	&	0.008	\\
	&		&	Histidine Degradation III	&	HAL	&	0.012	\\
	&		&	Heme Biosynthesis II	&	ALAS2	&	0.014	\\
	&		&	Glycogen Degradation III	&	MGAM	&	0.020	\\
	&		&	IL-10 Signaling	&	BLVRB,IL1R1	&	0.023	\\
	&		&	Granulocyte Adhesion and Diapedesis	&	C5AR1,IL1R1	&	0.034	\\

\midrule

\multirow{20}{*}{\centering Proteins}	&	\multirow{10}{*}{\centering Males}	&	Wound Healing Signaling Pathway	&	CERT1,HBEGF,PDGFA,PDGFB,SHC1	&	$<$0.001	\\
	&		&	PDGF Signaling	&	PDGFA,PDGFB,SHC1	&	$<$0.001	\\
	&		&	PPAR Signaling	&	PDGFA,PDGFB,SHC1	&	$<$0.001	\\
	&		&	Pulmonary Fibrosis Idiopathic Signaling Pathway	&	CERT1,PDGFA,PDGFB,THBS1	&	$<$0.001	\\
	&		&	PAK Signaling	&	PDGFA,PDGFB,SHC1	&	$<$0.001	\\
	&		&	Glioma Signaling	&	PDGFA,PDGFB,SHC1	&	$<$0.001	\\
	&		&	DHCR24 Signaling Pathway	&	APP,PDGFA,PDGFB	&	$<$0.001	\\
	&		&	Glioblastoma Multiforme Signaling	&	PDGFA,PDGFB,SHC1	&	$<$0.001	\\
	&		&	Regulation of the Epithelial Mesenchymal Transition by Growth Factors Pathway	&	PDGFA,PDGFB,SHC1	&	$<$0.001	\\
	&		&	Hepatic Fibrosis / Hepatic Stellate Cell Activation	&	CERT1,PDGFA,PDGFB	&	$<$0.001	\\

 \cmidrule{2-5}
									
	&	\multirow{10}{*}{\centering Females}	&	Wound Healing Signaling Pathway	&	CERT1,HBEGF,PDGFA,PDGFB,SHC1	&	$<$0.001	\\
	&		&	PDGF Signaling	&	PDGFA,PDGFB,SHC1	&	$<$0.001	\\
	&		&	PPAR Signaling	&	PDGFA,PDGFB,SHC1	&	$<$0.001	\\
	&		&	Pulmonary Fibrosis Idiopathic Signaling Pathway	&	CERT1,PDGFA,PDGFB,THBS1	&	$<$0.001	\\
	&		&	PAK Signaling	&	PDGFA,PDGFB,SHC1	&	$<$0.001	\\
	&		&	Glioma Signaling	&	PDGFA,PDGFB,SHC1	&	$<$0.001	\\
	&		&	DHCR24 Signaling Pathway	&	APP,PDGFA,PDGFB	&	$<$0.001	\\
	&		&	Glioblastoma Multiforme Signaling	&	PDGFA,PDGFB,SHC1	&	$<$0.001	\\
	&		&	Regulation of the Epithelial Mesenchymal Transition by Growth Factors Pathway	&	PDGFA,PDGFB,SHC1	&	$<$0.001	\\
	&		&	Hepatic Fibrosis / Hepatic Stellate Cell Activation	&	CERT1,PDGFA,PDGFB	&	$<$0.001 \\
 \bottomrule
\end{tabular}
\end{sidewaystable}